# Convolutional neural network with a hybrid loss function for fully automated segmentation of lymphoma lesions in FDG PET images

Fereshteh Yousefirizi*[a], Natalia Dubljevic[b], Shadab Ahamed [a,b], Ingrid Bloise[c], Claire Gowdy[d], Joo Hyun O[e], Youssef Farag[a], Rodrigue de Schaetzen[a], Patrick Martineau[c], Don Wilson[c], Carlos F. Uribe[c,f], Arman Rahmim[a,b,f]

[a]Department of Integrative Oncology, BC Cancer Research Institute, Vancouver, Canada;
[b]Department of Physics & Astronomy, University of British Columbia, Vancouver, BC, Canada;
[c]BC Cancer, Vancouver, Canada; [d]BC Children's Hospital, Vancouver, Canada; [e]Seoul St. Mary's Hospital, College of Medicine, The Catholic University of Korea, Seoul, Republic of Korea;
[f]Department of Radiology, University of British Columbia, Vancouver, Canada

## ABSTRACT

Segmentation of lymphoma lesions is challenging due to their varied sizes and locations in whole-body PET scans. In this work, we present a fully-automated segmentation technique using a multi-center dataset of diffuse large B-cell lymphoma (DLBCL) with heterogeneous characteristics. We utilized a dataset of [18F]FDG-PET scans (n=194) from two different imaging centers including cases with primary mediastinal large B-cell lymphoma (PMBCL) (n=104). Automated brain and bladder removal approaches were utilized as preprocessing steps, to tackle false positives caused by normal hypermetabolic uptake in these organs. Our segmentation model is a convolutional neural network (CNN), based on a 3D U-Net architecture that includes squeeze and excitation (SE) modules. Hybrid distribution, region, and boundary-based losses (Unified Focal and Mumford-Shah (MS)) were utilized that showed the best performance compared to other combinations (p<0.05). Cross-validation between different centers, DLBCL and PMBCL cases, and three random splits were applied on train/validation data. The ensemble of these six models achieved a Dice similarity coefficient (DSC) of $0.77 \pm 0.08$ and Hausdorff distance (HD) of $16.5 \pm 12.5$. Our 3D U-net model with SE modules for segmentation with hybrid loss performed significantly better (p<0.05) as compared to the 3D U-Net (without SE modules) using the same loss function (Unified Focal and MS loss) (DSC= $0.64 \pm 0.21$ and HD= $26.3 \pm 18.7$). Our model can facilitate a fully automated quantification pipeline in a multi-center context that opens the possibility for routine reporting of total metabolic tumor volume (TMTV) and other metrics shown useful for the management of lymphoma.

**Keywords:** Lymphoma, Convolutional neural network, PET, Segmentation, U-Net, Focal loss, hybrid loss

## 1. INTRODUCTION

Lymphoma is the cancer of the lymphatic system that can present in any of the over 500 lymph nodes and lymphatic organs such as the bone marrow and spleen. There are vast range of treatments from chemotherapy and radiation to newer immunotherapies and it is difficult to identify the patients who will not respond to therapy or who will acquire resistance to treatment (non-responder patients). With the widespread use of [18F]FDG PET imaging as the modality of choice for lymphoma staging and treatment response assessment, accurate lesion segmentation and tumor burden quantification from whole-body PET images is a crucial bottleneck towards improved prognostic pipelines[1] and treatment planning[2,3].

Total metabolic tumor volume (TMTV) measured from baseline [18F]FDG PET images of diffuse large B-cell lymphoma (DLBCL) and primary mediastinal large B-cell lymphoma (PMBCL) has strong prognostic value for progression-free and overall survival prediction. Among the predictive image-derived quantitative parameters; (e.g. total metabolic tumor volume (TMTV)) none is routinely reported in clinical practice. This is in part due to the time-consuming task of manual segmentation[4] and the fact that delineations are known to suffer from intra- and inter-observer and center variabilities[5,6]. As such, one possible solution to this problem is to develop a fully automated, generalizable and reproducible segmentation workflow.

Although, convolutional neural networks (CNNs) have shown good performances for medical image segmentation [7,8], specifically for PET segmentation[9-13], lymphoma lesion segmentation is a challenging task due to the varied number, size, site and shape of lesions, heterogeneity and different degrees of glucose metabolism[14-17]. The high rate of false positive is the other major challenge of lymphoma lesion segmentation caused by normal hypermetabolic organs (such as brain and bladder)[18]. Multi-center data are usually highly heterogeneous given different scanners, reconstruction algorithms, and post-processing steps across centers. In this study, we develop and validate a fully automated segmentation technique for lymphoma lesions in [18F]FDG-PET scans in the context of multi-center images that have lesions with different sizes and locations from diffuse large B cell (DLBCL) and its subtype i.e. primary mediastinal large B-cell lymphoma (PMBCL). Besides, the loss functions for training the CNNs has been shown to play an important role and affect the segmentation performance[19]. This emphasizes the idea that the specific design for loss functions should be considered for the challenging task of lymphoma lesion segmentation.

To this end, we have the following two goals in this study: 1) Exploring the capability of a 3D U-Net with squeeze and excitation (SE) modules for accurate and automated lymphoma lesion segmentation on PET scans from different centers without additional prior knowledge (such as bounding boxes) or post-processing steps. 2) Investigating the effect of using hybrid loss function consisting of region, boundary and distribution based losses on the semantic segmentation performance.

## 2. MATERIAL AND METHODS

**Data**

Table *1* summarizes the main characteristics of the data. The dataset of PET scans (n=194) were composed of DLBCL PET scans (n=90) from two different centers (BC Cancer (n=46) and South Korea (n=44)) and PMBCL PET scans (n=104) from BC Cancer. The ground truth volumes of interest (VOI) were delineated by experienced nuclear medicine physicians using a semi-automatic gradient-based segmentation (PETedge) from MIM (MIM Software, USA). This delineation method helped reduce inter-observer variability compared to manual delineations. Data-hungry supervised AI techniques require large number of delineated cases. To increase the diversity in lesion size and shape, we utilized scaling (with a random factor) and elastic deformations for data augmentation to help the model to learn the varied size and shape of the lesions

Table 1. Multi-center Dataset information from different lymphoma types

| Center | Lymphoma Type | Matrix size | Voxel spacing (mm$^3$) | Bed Duration (sec) | Average Injected Radioactivity (MBq) | Reconstruction Method | Scanner Models |
|---|---|---|---|---|---|---|---|
| BC Cancer, Canada | PMBCL | 168x168 (n=20) 192x192 (n=84) | 4.06x4.06x2 (n=20) 4.06x4.06x3.27 (n=119) | 120 (n=53) 150 (n=46) 180 (n=28) 210 (n=11) 240 (n=1) | 347.5±52.6 | OSEM 2D (3 It, 8 subs) | Siemens (1080) |
| BC Cancer, Canada | DLBCL | 192x192 | 3.65x3.65x3.27 (n=46) | 120 (n=23) 150 (n=16) 180 (n=9) 210 (n=2) | 335.9±50.8 | OSEM 3D No TOF with PSF(2 It, 32subs, 6.4 mm filter) | GE (Discovery D600 and D690) |
| St. Mary's Hospital, South Korea | Limited stage DLBCL | 168x168 (n=27) 192x192 (n=15) | 3.65x3.65x3.27 (n=15) 3.65x3.65x5 (n=27) | 90 (n=14) 108 (n=1) 120 (n=24) 150 (n=3) | 252.0±48.1 | OSEM3D No TOF and PSF(4 It, 8 subs) | GE (Discovery 710) |

### 2.1. Preprocessing steps

The segmentation performance can be affected by the variability of voxel sizes in the PET images (as shown in Table 1) since convolutional neural networks (CNNs) are not capable of interpreting the spatial dimensions with different scales[20]. We changed the PET images and their ground truth masks to the same resolution of 1 ×1×1 mm$^3$ with trilinear interpolation. We applied the linear interpolation since it does not decrease the segmentation performance compared to nearest neighbor and B-spline. To decrease the variabilities due to the varied intensity of PET images, we also applied Z-score normalization on each scan with the mean and standard deviation of non-zero voxels of body region.

Manual or automated removal of normal hypermetabolic organs is a common pre-processing step for whole-body PET image segmentation [21] that highly affects the performance of lesion segmentation[2]. As lesions in the bladder are rare in lymphoma (although there can be lesions very close to it), we removed these in the pre-processing step. The brain area was segmented automatically in the scans with brain area by applying thresholding and finding the biggest connected component in the upper part of the whole-body PET images. We also used a CNN for bladder segmentation that was previously trained on prostate-specific membrane antigen (PSMA) PET scans ([$^{18}$F]DCFPyL PET)[22]. The segmented bladder regions were removed from [$^{18}$F]FDG PET scans.

### 2.2. Segmentation Network Architecture

We utilized the 3D U-Net model (Figure 1) with residual layers, supplemented with SE normalization and learnable non-linear downsampling and upsampling branches previously proposed by Iantsen et al.[23]. The SE module gives weight to the feature maps, so that the network can emphasize its attention adaptively. The squeeze & excitation (SE) module is the architectural unit that is designed to increase the representational ability of the network by enabling dynamic channel-wise feature recalibration. SE module 'squeeze' along the spatial domain and 'excite' along the channels that helps the model to highlight the meaningful features and suppress the weak ones. CNNs with SE modules frequently achieve top performance, across various challenges e.g. ILSVRC 2017 image classification[24] and Head and Neck Tumor segmentation challenge (HECKTOR 2020)[23]. We used SE normalization layers with the fixed reduction ratio (r = 2) that controls the size of the bottleneck in SE normalization layers.

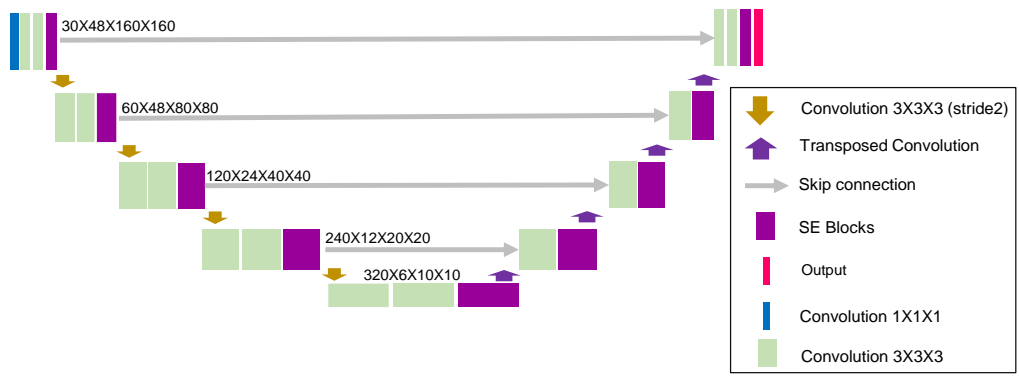

Figure 1: Proposed 3D U-net Network with residual blocks. The Squeeze and Excitation (SE) modules are shown in violet.

### 2.3. Training Process

The model was trained for 300 epochs using Adam optimizer on two NVIDIA Tesla V100 GPUs 16 GB with a batch size of 2. The cosine annealing schedule was applied to reduce the learning rate from $10^{-3}$ to $10^{-6}$ within every 25 epochs.

### 2.4. Hybrid Loss function

The loss function defines the optimization problem and directly affects the segmentation model convergence during training. The role of loss function is critical in deep learning pipeline. Our task is a semantic segmentation task that performs the pixel-level classification, in contrast to instance segmentation in which an additional class identification/prediction step with corresponding loss function is also required. The cross-entropy loss and Dice loss or a combination of them are mainly used for semantic segmentation. The different segmentation performance across different loss functions has been shown to emphasize the importance of loss function selection and how it affects the robustness and convergence of the segmentation model[19]. There are three main categories of loss functions that have been used for medical image segmentation: distribution-based losses (e.g. cross entropy, Focal loss[25]), region-based losses (e.g. Dice), boundary-based loss (e.g. Mumford-Shah[26]) and the combinations of them. The cross-entropy based losses have shown that are not a good reflection of segmentation quality and region based loss functions such as Dice loss are preferred for segmentation[27]. Some variations of the Dice losses have been proposed in this regard that are also capable of accounting the class imbalance in medical image segmentation tasks[28-30]. It has been shown that the best performance is usually observed with hybrid loss functions[19,31] e.g. the sum of cross entropy and Dice similarity coefficient (DSC) proposed by Taghanaki et al.[32] or the Unified Focal loss introduced by Yeung et al.[19] that combined Focal and Dice (the negative of DSC) loss.

In this study, we used a combination of distribution-based, region-based and boundary-based loss functions. Our proposed loss function incorporates the Unified Focal loss[19] and Mumford-Shah loss to take the advantage of boundary loss that optimizes boundaries rather than distributions or regions. The best model was selected based on the lowest validation loss. We used the hyperparameter of the loss functions as reported in the original studies and set the optimal values for the fair comparison (Table 2). The data were split into training, validation and test sets based on lymphoma types DLBCL cases from BC Cancer (DBC), PMBCL cases from BC Cancer (PBC), DLBCL cases from St. Mary's Hospital (DSM). We also used random training and validation splits.

The average results are provided for each evaluation metric across all the leave-one-out cross-validation. Three models were built by using the data from two parts for training and the data from the third part was held out for validation. Three other models were also fitted on random training/validation splits of the entire dataset. Predictions on the test set were produced by ensemble of the individual models. Three models also trained on random splits. The ensemble of these trained models (6 in total) was then applied on the test data.

## 3. RESULTS

The brain and bladder were segmented and removed prior to lymphoma lesion segmentation. On the original PSMA scans, a mean DSC of 0.911±0.096 was achieved and on our [$^{18}$F] FDG scans, a mean DSC of 0.867±0.10 was achieved for bladder segmentation. Two examples of bladder segmentation are shown in Figure 2. Our results were produced on the test set with the use of the ensemble of six models trained and validated on different splits of the training set.

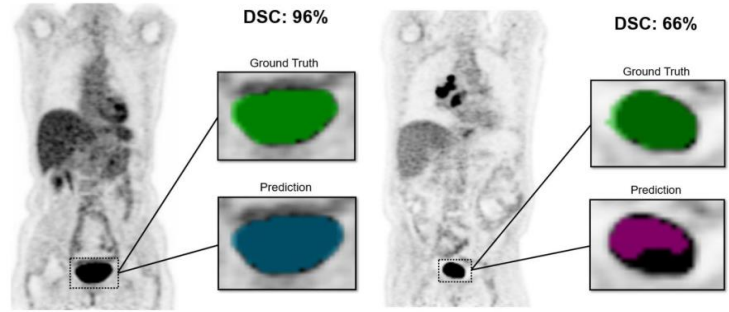

Figure 2. Shown are one of the best and one of the worst bladder segmentation on two different DLBCL patients. The Dice score was calculated in 3D, and a representative coronal slice chosen for visualization.

As shown in Table 2, our results revealed that 3D SE U-Net with only distribution-based (Focal) or region-based (soft Dice) losses had the lower performance and the hybrid loss (Focal + soft Dice) improved the segmentation results while using MS with Focal and Dice decreased the DSC a little but slightly improved HD. The Unified Focal has shown better performance compared to the other hybrid losses, but the hybrid loss function of Unified Focal and MS showed the best performance. The mean DSC (± standard deviation) of the segmentation model using the combination of Unified focal and MS loss are DSC= 0.77 ± 0.08 and the mean HD of 16.48 ±12.45 which are better as compared to the performance of 3D U-net without SE modules using the same hybrid loss function (DSC= 0.64 ± 0.21 and HD= 26.34 ± 18.67). To test for statistical significance, we used the Wilcoxon rank-sum test for statistical significance ($p < 0.05$). Our 3D SE U-Net segmentation model with our proposed hybrid loss missed some small lesions in DLBCL cases (6 lesions in n=6 DLBCL cases) and PMBCL (one lesion in n=1 case); an example is shown in Figure 3.

Table 2. The effect of loss function on quantitative segmentation results by our proposed method on test data

| Loss function | Hyper-parameters | Dice | HD |
|---|---|---|---|
| Soft Dice | - | 0.68 ± 0.21 | 24.6 ± 27.3 |
| Focal | α=0.25, γ=2 | 0.72 ± 0.24 | 28.7 ± 19.2 |
| Focal + soft Dice | α=0.25, γ=2 | 0.75 ± 0.16 | 31.2 ± 21.9 |
| Focal + soft Dice + MS | α=0.3, γ=2, β=10$^{-6}$ | 0.74 ± 0.18 | 29.6 ± 22.7 |
| Unified Focal | λ=0.5, δ=0.6, γ=0.5 | 0.76 ± 0.10 | 17.5 ± 13.10 |
| Unified Focal + MS | λ=0.5, δ=0.6, γ=0.5, β=10$^{-7}$ | 0.77 ± 0.08 | 16.5 ±12.5 |

The best segmentation performance in terms of Dice and Hausdorff distance (HD) metrics was received for the 'PBC' data with 104 patients (DSC=0.78 ±0.5, HD=14.4 ± 8.1) and the poorest performance was achieved for the 'DSM' with 42 patients (DSC=0.69 ± 0.7, HD=19.5 ± 19.8). The segmentation performance of 3D SE U-Net with hybrid loss was noticeably decreased when the bladder was not removed, and only slightly decreased when the brain was not removed (Table 3). In order to reduce the effect of normal organ removals on each other the comparisons were done using the model with just one organ removal.

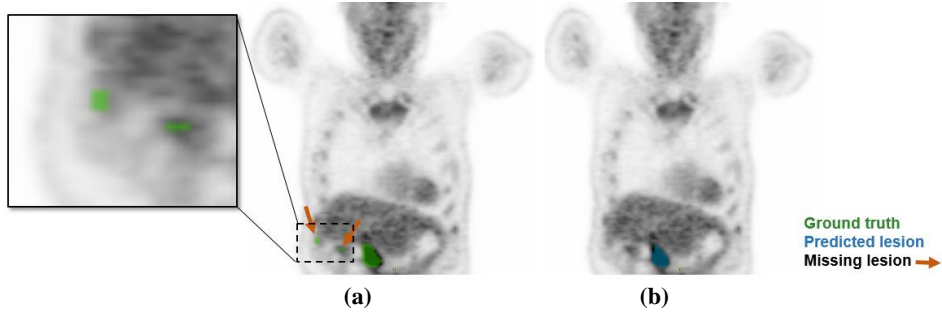

Figure 3. The missed small lesions by our 3D SE U-Net in a DLBCL case (a) the ground truth with two small lesions that have been shown in the box, (b) the predicted mask by our segmentation model of the same patient. The two small lesions are missed by our segmentation approach.

Table 3. The effect of normal organ removal on the segmentation results

| Normal organ removal (Pre-processing) | DSC(with) | DSC (without) | HD(with) | HD (without) |
|---|---|---|---|---|
| Bladder (no brain removal) | 0.76 ± 0.23 | 0.73 ± 0.18 | 16.9 ±13.6 | 17.5 ± 12.3 |
| Brain (no bladder removal) | 0.74 ± 0.19 | 0.73 ± 0.18 | 18.2 ±11.2 | 17.5 ± 12.3 |

## 4. DISCUSSION & CONCLUSION

Neither manual brain and bladder removal nor bounding boxes or any center specific standardization were required for the challenging task of lymphoma lesion segmentation with our proposed 3D SE U-Net model with hybrid loss function. The performance of the proposed segmentation model without the bladder removal is reduced but without brain removal, the results were only slightly changed. This is due to the fact that the brain region was mostly not captured in lymphoma scans. On the other hand, considering the possible metastasis in brain area for lymphoma cancer, brain removal may not be a certain pre-processing step. However, there is no brain metastasis in the patient of our dataset. Using bi-modal PET and CT inputs for segmentation model was also considered that showed no specific increase in the segmentation performance in most of the DLBCL and PMBCL cases, although we assumed by using CT data the more accurate boundary results would be achieved.

The best segmentation performance of our model was observed using the proposed hybrid loss function that incorporates the distribution, region and boundary-based loss functions (Unified Focal loss and Mumford-Shah). Volume-aware loss functions can help optimize the training of the segmentation model to extract the small lesions with higher performance and decrease the missed small lesion rate, that we will consider it in our future works.

We developed and validated a fully-automated pipeline segmenting lymphoma lesions with varied type, size, sites, and location from different institutions. This segmentation scheme is aimed to facilitate automated prognostic workflows, including routine, automated quantification of total metabolic tumor volume (TMTV) as a known prognostic imaging biomarker. Our proposed 3D CNN model (3D U-net with SE modules) can facilitate fully-automated quantification pipeline in a multi-center context. Overall, the differences with the three categories of data, DLBCL cases from BC Cancer (DBC), PMBCL cases from BC Cancer (PBC), DLBCL cases from St. Mary's Hospital (DSM) implied the segmentation model is robust to the different centers and no center specific data standardization was needed. The lower performance on

DLBCL data from DSM is due to the smaller lesions of limited stage patients from this specific center. The insignificant differences in the average results between the leave-one-out and random split cross validations supports this finding.

## ACKNOWLEDGEMENT

This research was supported in part through computational resources and services provided by Microsoft and by Canadian Institutes of Health Research (CIHR) Project Grant PJT-173231.